\documentclass[condensedmatter,article,accept,oneauthor,
LaTex,10pt,a4paper]{mdpi} 

\firstpage{1} 
\articlenumber{x}
\doinum{10.3390/------}
\pubvolume{2}
\pubyear{2017}
\copyrightyear{2017}
\externaleditor{Academic Editor: Andrea Perali and Alessandro Ricci}
\history{{Received: 23 April 2017; Accepted: 7 June 2017; Published: date}}

\usepackage[utf8]{inputenc}
\usepackage[T1]{fontenc}
\usepackage{amsmath,amssymb,amsthm}
\usepackage{mathrsfs}  
\usepackage{bbold}
\usepackage{epsfig}
\usepackage{graphicx}

\newcommand{\beq}{\begin{equation}}
\newcommand{\eeq}{\end{equation}}
\newcommand{\beqa}{\begin{eqnarray}}
\newcommand{\eeqa}{\end{eqnarray}}

\Title{Goldstone and Higgs Hydrodynamics in the BCS{--}BEC Crossover}

\Author{Luca Salasnich $^{1,2}$}

\address{%
$^{1}$ \quad Dipartimento di Fisica e Astronomia ``Galileo Galilei'' 
and CNISM, Universit\`a di Padova, Via Marzolo 8, 35131 Padova, Italy; 
luca.salasnich@unipd.it
\\
$^{2}$ \quad Istituto Nazionale di Ottica (INO) del Consiglio Nazionale delle 
Ricerche (CNR), Via Nello Carrara 1, 50019~Sesto Fiorentino, Italy
}

\corres{Correspondence: luca.salasnich@unipd.it; Tel.: +39-049-8277132}

\abstract{We discuss the derivation of a low-energy 
effective field theory of phase (Goldstone) and 
amplitude (Higgs) modes of the pairing 
field from a microscopic theory of attractive fermions. 
The~coupled equations for Goldstone and Higgs fields 
are critically analyzed in the Bardeen--Cooper--Schrieffer (BCS){-}to {-}
Bose--Einstein condensate (BEC) crossover---both in three spatial dimensions 
and in two spatial dimensions. 
The crucial role of pair fluctuations is investigated, and the 
beyond-mean-field Gaussian theory of the BCS--BEC crossover 
is compared with available experimental data of the 
two-dimensional ultracold Fermi superfluid.}


\begin{document}

\section{Introduction}

An important achievement in the physics of ultracold atoms 
was the experimental realization of the crossover from the 
Bardeen--Cooper--Schrieffer (BCS) superfluid phase of loosely-bound pairs 
of fermions to the Bose--Einstein condensate (BEC) of tightly-bound composite 
bosons \cite{greiner2003,chin2004}. Superfluid fermions in the  {BCS--BEC} 
~crossover are an ideal platform to investigate macroscopic 
quantum phenomena. For instance, macroscopic quantum tunneling 
and quantum self-trapping \cite{smerzi,sala1999,sala-mqt} have recently been 
observed across the BCS--BEC crossover \cite{roati}. 

Very recently, 
the BCS--BEC crossover has also been realized in quasi two-dimensional (2D) 
configurations \cite{makhalov,murthy,fenech,boettcher}. 
Beyond-mean-field theoretical investigations of 2D Fermi gases 
in the full BCS--BEC crossover have been carried out 
both at zero and finite temperature 
\cite{sala2013,sala2015,cin2015,cin2015b,sala2016,sala-review,cin2017}. 
These 2D results have clearly shown that 
contrary to the 3D case, mean-field theories are completely unreliable 
for the study of strongly-interacting superfluid fermions in two dimensions 
because of the huge increase of quantum fluctuations. 
The 2D BCS--BEC crossover is also interesting
for high-T$_{c}$ superconductivity where the phase diagram of 
cuprate superconductors can be interpreted in terms of a BCS--BEC crossover 
as doping is varied. The critical temperature T$_c$ has a wide fluctuation 
region with pseudo-gap effects not yet fully understood 
\cite{larkin-varlamov,levin}. 
Moreover, it has been suggested that iron-based superconductors 
have composite superconductivity, consisting of strong-coupling BEC in 
the electron band and weak-coupling BCS-like superconductivity in the 
hole band \cite{okazaji}. 

This year, we have used renormalization-group techniques 
to calculate the Berezinskii--Kosterlitz--Thouless (BKT) critical temperature 
\cite{berezinskii,kosterlitz}, 
taking explicitly into account the formation of quantized 
vortices \cite{sala2017}. The presence of quantized vortices and antivortices 
renormalizes the superfluid density of the system as the temperature 
increases, and the renormalized superfluid density jumps from 
a finite value to zero as the temperature reaches the BKT critical 
temperature \cite{nelson}. In this way, we have also 
obtained the BKT critical temperature in the full 2D BCS--BEC 
crossover for the uniform Fermi superfluid \cite{sala2017}. 
Quantized vortices in superfluids are a peculiar 
consequence of the existence of an underlying compact real field, whose 
spatial gradient determines the local superfuid velocity of the system 
\cite{nagaosa,altland,stoof}. This compact real field---the so-called Goldstone field---is the phase angle of the 
complex bosonic field which in the case of 
attractive fermions describes strongly-correlated Cooper pairs of fermions with 
opposite spins. The amplitude of this complex pairing field---sometimes also called  { a} Higgs field \cite{varma}---is decoupled from 
the angular Goldstone field only in the deep BCS regime where 
the particle--hole symmetry is almost preserved.  

In this paper, we review the low-energy effective field theory 
of Goldstone and Higgs modes derived from the microscopic theory 
of paired fermions. At zero temperature, we show that the velocity 
of the Goldstone mode obtained with and without amplitude 
fluctuations (in 2D but also in 3D) has a quite different 
behavior \cite{sala2013}. We find that amplitude fluctuations 
are necessary to identify 
the velocity of the Goldstone mode with the first sound velocity 
one gets from the mean-field equation of state by using familiar 
thermodynamics relationships \cite{sala2013}. 
Finally, we show that in the 2D case, the first sound velocity 
obtained from the beyond-mean-field equation of state is quite 
different with respect to the mean-field one---in particular 
in the BEC side of the BCS--BEC crossover \cite{sala2016}. 
For the sake of completeness, we also report our beyond-mean-field 
calculation \cite{cin2015,cin2015b,sala2016,sala-review} 
of the zero-temperature pressure of the Fermi superfluid 
in the 2D BCS--BEC crossover: these  {theoretical} ~results are in very good 
agreement with the available experimental data \cite{makhalov}. 

\section{Functional Integration for the BCS--BEC Crossover} 

We consider a D-dimensional system of two-spin-component fermions 
interacting through 
an attractive $s$-wave contact potential, contained in a volume $V$, 
at fixed chemical potential $\mu$ and temperature $T$. 
Within the path integral formalism, the partition 
function of the system can be written as \cite{nagaosa,altland,stoof}
\beq
\mathcal{Z} = \int \mathcal{D} \psi_\sigma \mathcal{D} \bar{\psi}_\sigma 
\ e^{- \frac{1}{\hbar} \int_0^{\hbar \beta} \mathrm{d} \tau \int_{V} 
\mathrm{d}^D{\bf r} \ \mathscr{L}}
\label{eq:z}
\eeq
where $\psi_{\sigma}({\bf r},\tau)$, $\bar{\psi}_{\sigma}({\bf r},\tau)$ 
are complex Grassmann fields ($\sigma=\uparrow,\downarrow$), 
$\beta=1/k_B T$, with $k_B$ is Boltzmann constant, and the Euclidean 
Lagrangian density reads 
\beq
\mathscr{L} = \bar{\psi}_{\sigma} \left[ \hbar \partial_{\tau} 
- \frac{\hbar^2}{2m}\nabla^2 - \mu \right] \psi_{\sigma} 
+ g \, \bar{\psi}_{\uparrow} \, \bar{\psi}_{\downarrow} 
\, \psi_{\downarrow} \, \psi_{\uparrow} \; .
\label{eq:l}
\eeq

Here $g$ is the attractive interaction strength ($g<0$) of the s-wave coupling 
between fermions with opposite spin $\sigma$. 
Looking for analytical solutions, the quartic interaction 
cannot be treated exactly. The 
Hubbard--Stratonovich transformation \cite{nagaosa,altland,stoof} 
introduces an additional auxiliary pairing field $\Delta({\bf r},\tau)$ 
corresponding to a 
Cooper pair, decoupling the quartic interaction. The transformation 
is based on the following identity
\beq
e^{- \frac{1}{\hbar} \int_0^{\hbar \beta} \mathrm{d} \tau \int_{V} 
\mathrm{d}^D{\bf r} g 
\, \bar{\psi}_{\uparrow} \, \bar{\psi}_{\downarrow} 
\, \psi_{\downarrow} \, \psi_{\uparrow}} = \int \mathcal{D} \Delta 
\mathcal{D} 
\bar{\Delta} \ e^{\frac{1}{\hbar} 
\int_0^{\hbar \beta} 
\mathrm{d} \tau \int_{V} \mathrm{d}^D{\bf r} \left( \frac{|\Delta|^2}{g} + 
\bar{\Delta} \psi_\downarrow \psi_\uparrow + \Delta  \bar{\psi}_\uparrow 
\bar{\psi}_\downarrow \right) } \; . 
\eeq

The partition function can then be written as 
\beq
\mathcal{Z} = \int \mathcal{D} \psi_\sigma \mathcal{D} \bar{\psi}_\sigma 
\mathcal{D} \Delta \mathcal{D} \bar{\Delta} \ e^{- \frac{1}{\hbar} 
\int_0^{\hbar \beta} 
\mathrm{d} \tau \int_{V} \mathrm{d}^D{\bf r} \ \mathscr{L}_e }
\label{eq:z2}
\eeq
with the following Lagrangian density
\beq
\mathscr{L}_e =
\bar{\psi}_{\sigma} \left[  \hbar \partial_{\tau} 
- {\hbar^2\over 2m}\nabla^2 - \mu \right] \psi_{\sigma} 
+ \bar{\Delta} \, \psi_{\downarrow} \, \psi_{\uparrow} 
+ \Delta \bar{\psi}_{\uparrow} \, \bar{\psi}_{\downarrow} 
- {|\Delta|^2\over g} \; .
\label{ltilde}
\eeq

The integration over the fermionic fields 
$\psi_{\sigma}({\bf r},\tau)$ and $\bar{\psi}_{\sigma}({\bf r},\tau)$ 
can now be carried out exactly,~obtaining 
\beq
\mathcal{Z} = \int \mathcal{D} \Delta \mathcal{D} \bar{\Delta} e^{- \frac{1}
{\hbar} \int_0^{\hbar \beta} \mathrm{d} \tau \int_{V} \mathrm{d}^D{\bf r} 
\ \left( - \ln ( - \mathbb{G}^{-1} \right) - {|\Delta|^2 \over g})}
\label{eq:z3}
\eeq
with $\mathbb{G}^{-1}$ the inverse Green's function, given by 
\beq
- \mathbb{G}^{-1} = \begin{pmatrix} \hbar \partial_\tau - \frac{\hbar^2}{2m} 
\nabla^2 - \mu & \Delta(\mathbf{r},\tau) \\
\bar{\Delta}(\mathbf{r},\tau) & \hbar \partial_\tau + 
\frac{\hbar^2}{2m} \nabla^2 +\mu \end{pmatrix} \; .
\eeq

To investigate effects of quantum and thermal fluctuations of the gap 
field $\Delta({\bf r},t)$ around its mean-field value $\Delta_0$, we set 
\beq 
\Delta({\bf r},\tau) = \Delta_0 +  \eta(\mathbf{r},\tau) \; , 
\eeq
where $\eta({\bf r},\tau)$ is the complex field of pairing fluctuations. 
In this way, the inverse Green function $\mathbb{G}^{-1}$ 
is decomposed in a mean-field component 
$- \mathbb{G}^{-1}_0$, where the pairing field $\Delta$ is replaced by its 
uniform and constant saddle point value, plus a fluctuation part $\mathbb{F}$:
\beq
- \mathbb{G}^{-1} =  - \mathbb{G}^{-1}_{0} + \mathbb{F} = \begin{pmatrix} 
\hbar \partial_\tau - \frac{\hbar^2}{2m} \nabla^2 - \mu & \Delta_0 \\
\Delta_0 & \hbar \partial_\tau + \frac{\hbar^2}{2m} \nabla^2 +\mu 
\end{pmatrix} + \begin{pmatrix} 0 & \eta(\mathbf{r},\tau) \\
\bar{\eta}(\mathbf{r},\tau) & 0 \end{pmatrix}
\label{eq:expansion}
\eeq 

\subsection{Loop Expansion and Gaussian Approximation}

The logarithm appearing in Equation (\ref{eq:z3}) can be written 
as \cite{Tempere:2012cm}:
\beq
\ln \left( - \mathbb{G}^{-1} \right) = \ln \left(  - \mathbb{G}^{-1}_{0} 
\left( \mathbb{I} - \mathbb{G}_{0} \mathbb{F} \right) \right) = 
\ln \left(  - \mathbb{G}^{-1}_{0} \right) + \ln \left( \mathbb{I} - 
\mathbb{G}_{0} \mathbb{F} \right) \; . 
\label{eq:f1}
\eeq

The Gaussian (one-loop) approximation consists of the following 
expansion for the second term in the right-hand-side of Equation (\ref{eq:f1}):
\beq
\ln \left( \mathbb{I} - \mathbb{G}_{0} \mathbb{F} \right) = 
- \sum_{m=1}^\infty \frac{\left( \mathbb{G}_{0} \mathbb{F} \right)^m}{m} 
\simeq - \mathbb{G}_{0} \mathbb{F}  - \frac{1}{2} \mathbb{G}_{0} 
\mathbb{F}  \mathbb{G}_{0} \mathbb{F} \; .
\eeq

Within this Gaussian approximation, the partition function reads 
\beq
\mathcal{Z} \simeq \mathcal{Z}_{mf}  \ \mathcal{Z}_g \; , 
\eeq
where 
\beq 
\mathcal{Z}_{mf} = e^{- \frac{1}
{\hbar} \int_0^{\hbar \beta} \mathrm{d} \tau \int_{V} \mathrm{d}^D{\bf r} 
\ \left( - \ln ( - \mathbb{G}^{-1}_0 \right) - {|\Delta_0|^2 \over g})}
\eeq 
is the mean-field partition function and 
\beq 
\mathcal{Z}_g = 
\int \mathcal{D} \eta \mathcal{D} \bar{\eta} e^{- \frac{1}{\hbar} 
S_g [\eta,\bar{\eta}]}
\label{eq:z4}
\eeq
is the Gaussian partition function characterized by the following Gaussian 
action:
\beq
S_{g} [\eta,\bar{\eta}] = {1\over 2} \sum_{\mathbf{q}, m} 
({\bar\eta}(\mathbf{q}, \mathrm{i} \Omega_m),\eta(-\mathbf{q}, - \mathrm{i} 
\Omega_m)) \ \mathbb{M} (\mathbf{q}, \mathrm{i} \Omega_m) \left(
\begin{array}{c}
\eta(\mathbf{q}, \mathrm{i} \Omega_m) \\ 
{\bar\eta}(-\mathbf{q}, - \mathrm{i} \Omega_m) 
\end{array}
\right) \; . 
\label{g-action}
\eeq

In this formula, we have introduced the Fourier transform 
of the fluctuation fields and the bosonic Matsubara frequencies $\Omega_m = 
2 m \pi/ \beta$. 
The matrix elements of the inverse pair fluctuation propagator 
$\mathbb{M}$ are given by \cite{Tempere:2012cm} 
\begin{multline}
\mathbb{M}_{11}(\mathbf{q},\mathrm{i} \Omega_m) = 
- \frac{1}{\bf g} + \sum_\mathbf{k} 
\frac{\tanh(\beta E_{sp}({\bf k})/2) }{2 E_{sp}({\bf k})} \times \\
\times \left[ \frac{(\mathrm{i} \Omega_m - E_{sp}({\bf k}) + 
{\hbar^2({\bf k}+{\bf q})^2\over 2m} - \mu 
)(E_{sp}({\bf k}) + {\hbar^2k^2\over 2m} - \mu )}
{(\mathrm{i} \Omega_m - E_{sp}({\bf k}) + E_{sp}({\bf k}+{\bf q})) 
(\mathrm{i} \Omega_m - E_{sp}({\bf k}) - E_{sp}({\bf k}+{\bf q}))} 
\right. \\
\left. - \frac{(\mathrm{i} \Omega_m + E_{sp}({\bf k}) + 
{\hbar^2({\bf k}+{\bf q})^2\over 2m} - \mu )
(E_{sp}({\bf k}) - {\hbar^2k^2\over 2m} + \mu)}
{(\mathrm{i} \Omega_m + E_{sp}({\bf k}) - E_{sp}({\bf k}+{\bf q})) 
(\mathrm{i} \Omega_m + E_{sp}({\bf k}) + E_{sp}({\bf k}+{\bf q}))} \right] \;
\end{multline}
and
\begin{multline}
\mathbb{M}_{12} (\mathbf{q},\mathrm{i} \Omega_m) = - \Delta_0^2 \sum_{\bf k} 
\frac{\tanh(\beta E_{sp}({\bf k})/2) }{2 E_{sp}({\bf k})} \times \\
\times \left[ \frac{1}{(\mathrm{i} \Omega_m - E_{sp}({\bf k}) + 
E_{sp}({\bf k}+{\bf q})) (\mathrm{i} \Omega_m - E_{sp}({\bf k}) -
E_{sp}({\bf k}+{\bf q}))}  \right. \\
\left. + 
\frac{1}
{(\mathrm{i} \Omega_m + E_{sp}({\bf k}) - E_{sp}({\bf k}+{\bf q}) ) 
(\mathrm{i} \Omega_m + E_{sp}({\bf k}) + E_{sp}({\bf k}+{\bf q})  ) }
\right] \; ,  
\end{multline}
where 
\beq 
E_{sp}({\bf k}) = 
\sqrt{\left({\hbar^2k^2\over 2m}-\mu\right)^2+\Delta_0^2}
\eeq
is the spectrum of fermionic single-particle excitations. 

\subsection{Beyond-Mean-Field Grand Potential}

The thermodynamic grand potential $\Omega$ of the fermionic superfluid 
is given by 
\beq 
\Omega = -{1\over \beta}  \ln{\left( {\cal Z} \right)} 
\eeq

At the Gaussian one-loop level, one gets 
\beq 
\Omega \simeq - {1\over \beta} 
\ln{\left( {\cal Z}_{mf} {\cal Z}_g \right)} = \Omega_{mf} + \Omega_{g} \; ,  
\eeq
where the mean-field grand potential reads \cite{nagaosa,altland,stoof}
\beq
\Omega_{mf} = - {\Delta_0^2\over g} V 
+ \sum_{\bf k} \left( {\hbar^2k^2\over 2m} - \mu - E_{sp}({\bf k}) 
- {2\over \beta } \ln{(1+e^{-\beta\, E_{sp}({\bf k})})} \right) \; .  
\eeq

The Gaussian grand potential is instead  
\beq
\Omega_{g} = {1\over 2\beta} \sum_{{\bf q},m} 
\ln{\mbox{det}(\mathbb{M}({\bf q},i\Omega_m))} \; .   
\label{goduria}
\eeq

The sum over Matsubara frequencies is quite complicated, and it 
does not give a simple expression. An approximate formula which is valid in the BEC regime of the crossover \cite{griffin}
is the following:
\beq
\Omega_{g} \simeq {1\over 2} 
\sum_{\bf q} E_{col}({\bf q}) + {1\over \beta }
\sum_{\bf q} \ln{(1- e^{-\beta\, E_{col}({\bf q})})} \; ,   
\eeq
where 
\beq 
E_{col}({\bf q}) = \hbar \ \omega({\bf q}) 
\eeq
is the spectrum of bosonic collective excitations with $\omega({\bf q})$ 
derived from 
\beq 
\mbox{det}(\mathbb{M}({\bf q}, \omega)) = 0 \; . 
\eeq

In the Gaussian pair fluctuation (GPF) approach \cite{drumm}, given  
the grand potential 
\beq 
\Omega(\mu,V,T,\Delta_0) = \Omega_{mf}(\mu,V,T,\Delta_0) + 
\Omega_g(\mu,V,T,\Delta_0) \; , 
\label{god-or-odd}
\eeq
the energy gap $\Delta_0$ is obtained from the mean-field gap equation 
\beq 
{\partial \Omega_{mf}(\mu,V,T,\Delta_0) \over \partial \Delta_0} = 0 \; . 
\eeq

The number density $n$ is instead obtained from the beyond-mean-field 
number equation 
\beq 
n = - {1\over V} {\partial \Omega(\mu,V,T,\Delta_0(\mu,T)) 
\over \partial \mu} \;,
\label{number}
\eeq
taking into account the gap equation (i.e., that $\Delta_0$ 
depends on $\mu$ and $T$: $\Delta_0(\mu,T)$). 
Notice that the Nozieres and Schmitt--Rink 
approach {es} \cite{nsr}  {are} quite similar, but in the number equation one 
forgets that $\Delta_0$ depends on $\mu$. 

\section{Low-Energy Gaussian Action} 

We have seen that the analytical form of the inverse pair 
fluctuation propagator $\mathbb{M}({\bf q},\omega)$ (with~$\omega =~i \Omega_m$) is quite complicated, and 
one can find  its matrix elements numerically. Here~we use a series expansion 
of $\mathbb{M}({\bf q},\omega)$ up to the second order in ${\bf q}$ 
and $\omega$ \cite{moit,moit2,schakel}. 
Moreover, we decompose the fluctuation field as follows: 
\beq 
\eta({\bf r},\tau) =  \sigma({\bf r},\tau) + 
i \ \Delta_0 \ \theta({\bf r},\tau)  \; , 
\eeq
where $\sigma({\bf r},\tau)$ and $\theta({\bf r},\tau)$ are real 
and can be identified at the lowest order with amplitude 
and phase fluctuations, respectively \cite{moit,moit2,schakel}. 
In other words, $\sigma({\bf r},\tau)$ 
is the Higgs field and $\theta({\bf r},\tau)$ is 
the Goldstone field \cite{varma}. 

In this way, after some calculations \cite{moit2,schakel,sala2013}, 
one finds the low-energy real-time Gaussian action derived from 
the Euclidean Gaussian action (\ref{g-action}) 
\beq 
S_{\theta\sigma} = \int dt \int_V d^D{\bf r} \left\{  
- {J\over 2} (\nabla \theta)^2 
+ {K_{\theta \theta}\over 2} 
\left({\partial \theta \over \partial t}\right)^2 
- { K_{\sigma\sigma}\over 2} \sigma^2 - K_{\theta\sigma} \ 
{\partial \theta\over \partial t} \ \sigma \right\}\;, 
\label{low-action}
\eeq
where $t=-i\tau$ is the real time. At zero temperature, 
the coefficients $J$, $K_{\theta\theta}$, 
$K_{\sigma\sigma}$, and $k_{\theta\sigma}$ are related 
to the partial derivatives of the zero-temperature 
mean-field grand potential $\Omega_{mf}(\mu,V,T=0,\Delta_0)$ 
\cite{sala2013,schakel}. In particular, 
\beq 
J = - {\hbar^2\over 4m} {1\over V} {\partial \Omega_{mf}\over \partial \mu} 
\label{stiff}
\eeq 
is the phase stiffness, 
\beq 
K_{\theta\theta} = {\hbar^2\over m} {1\over V} 
{\partial^2 \Omega_{mf}\over \partial \mu^2} 
\eeq
is the phase--phase susceptibility, 
\beq 
K_{\sigma\sigma} = {\hbar^2\over 2m} {1\over V} 
{\partial^2 \Omega_{mf}\over \partial \Delta_0^2} 
\eeq
is the amplitude--amplitude susceptibility, and 
\beq 
K_{\theta\sigma} = {\hbar^2\over m} {1\over V} 
{\partial^2 \Omega_{mf}\over \partial \mu \partial \Delta_0} 
\eeq
is the phase--amplitude susceptibility, that is the Goldstone-Higgs 
coupling constant. Equation (\ref{low-action}) is 
practically the same action functional derived in \cite{rivers2011,rivers2013} 
by using a two-channel model model for the BCS--BEC crossover. 

The Euler--Lagrange equations for the Goldstone field $\theta({\bf r},t)$ 
and Higgs field $\sigma({\bf r},t)$ obtained 
from (\ref{low-action}) are 
\beqa 
\left[ K_{\theta\theta} 
{\partial^2 \over \partial t^2} - J \ \nabla^2\right] \theta = 
K_{\theta\sigma} \ {\partial \sigma \over \partial t} \; , 
\label{pio1}
\\
K_{\sigma\sigma} \sigma  = - 
K_{\theta\sigma} {\partial \theta\over \partial t} \; . 
\label{pio2}
\eeqa

Calculating the time derivative of $\sigma({\bf r},t)$ from Equation (\ref{pio2}), 
one finds 
\beq 
{\partial \sigma \over \partial t} = - 
{K_{\theta\sigma}\over K_{\sigma\sigma}} 
{\partial^2 \theta\over \partial t^2} \; . 
\eeq

Inserting this result in Equation (\ref{pio1}), we finally get the 
d'Alambert equation of waves
\beq 
\left[ 
\left( K_{\theta\theta} +  {K_{\theta\sigma}^2\over K_{\sigma\sigma}}\right) 
{\partial^2 \over \partial t^2} - J \ \nabla^2\right] \theta = 0 \; ,  
\label{euler}
\eeq
which admits the generic solution 
\beq 
\theta({\bf r},t) = \theta_{0} \ \sin({\bf k}\cdot {\bf r} - \omega_k \ t 
+ \phi_0) 
\eeq
with the dispersion relation 
\beq 
\omega_k = c_s \ k 
\eeq
and 
\beq 
c_s = \sqrt{J\over K} 
\label{suono}
\eeq
the velocity of propagation of the Goldstone mode, where 
\beq 
K = {K_{\theta\theta} K_{\sigma\sigma}+ K_{\theta\sigma}^2 
\over K_{\sigma\sigma}} 
\label{kappa}
\eeq  
is the effective susceptibility. Notice that only if 
the amplitude fluctuations are negligible 
(i.e., $\sigma \simeq 0$); from Equation (\ref{pio2}), it 
follows $K_{\theta\sigma} \simeq 0$ and consequently 
$K \simeq K_{\theta\theta}$.  

In the 3D BCS--BEC crossover, the interaction strength $g$ is usually 
written in terms of the to 3D s-wave scattering length $a$ \cite{stoof}
\beq 
{1\over g} = {m\over 4\pi \hbar^2 a} + {1\over V} \sum_{\bf k} 
{1\over {\hbar^2k^2\over m}} \; . 
\label{divergo3d}
\eeq

Instead, in the 2D BCS--BEC crossover, 
the interaction strength $g$ is usually related to the 
binding energy of Cooper pairs by \cite{sala-review}
\beq 
{1\over g} = {1\over V} \sum_{\bf k} {1\over {\hbar^2k^2\over m}+ 
\epsilon_B} \; . 
\label{divergo2d}
\eeq 

In fact, contrary to the 3D case, 2D realistic interatomic 
attractive potentials  always have a bound state. Both 
Equations (\ref{divergo3d}) and (\ref{divergo2d}) are ultraviolet divergent,  
but they exactly compensate the divergence of 
the mean-field grand potential $\Omega_{mg}$, which depends 
of the bare interaction strength~$g$~\cite{sala-review}.

Figure 1 shows that taking into 
account only phase fluctuations (i.e., setting $K=K_{\theta\theta}$) 
leads to a quite different behaviour of the velocity $c_s$ 
from that obtained by considering both phase and amplitude fluctuations 
\cite{sala2013}. In the upper panel, there is the behaviour of $c_s$ in 
the 2D BCS--BEC crossover, while in the lower panel there are the results 
for the 3D BCS--BEC crossover. 
Only in the deep BCS regime (left~side of the two panels of Figure 1) 
do the two approaches give the same results, 
while the phase-only sound velocity diverges 
in the BEC regime (right side of panels). Thus, the effect 
of Goldstone--Higgs coupling $K_{\theta\sigma}$ is crucial in the study of 
the BCS--BEC crossover. It is important to stress that 
the zero-temperature results of Figure 1 are obtained 
adopting the mean-field number equation;~i.e., 
\beq 
n = - {1\over V} {\partial \Omega_{mf}(\mu,V,T=0,\Delta_0(\mu,T=0)) 
\over \partial \mu} \; .  
\label{number-mf}
\eeq

We used Equation (\ref{number-mf}) instead of Equation (\ref{number}) 
in the calculation of $c_s$ in order to have, from Equation (\ref{stiff}), 
\beq 
J = {\hbar^2\over 4m} n 
\eeq
which is the expected phase stiffness at zero temperature, 
where the total density $n$ should be equal to the superfluid density $n_s$. 
In this way, one also satisfies the compressibility sum rule \cite{levin2}. 

\begin{figure}[H]
\centerline{\epsfig{file=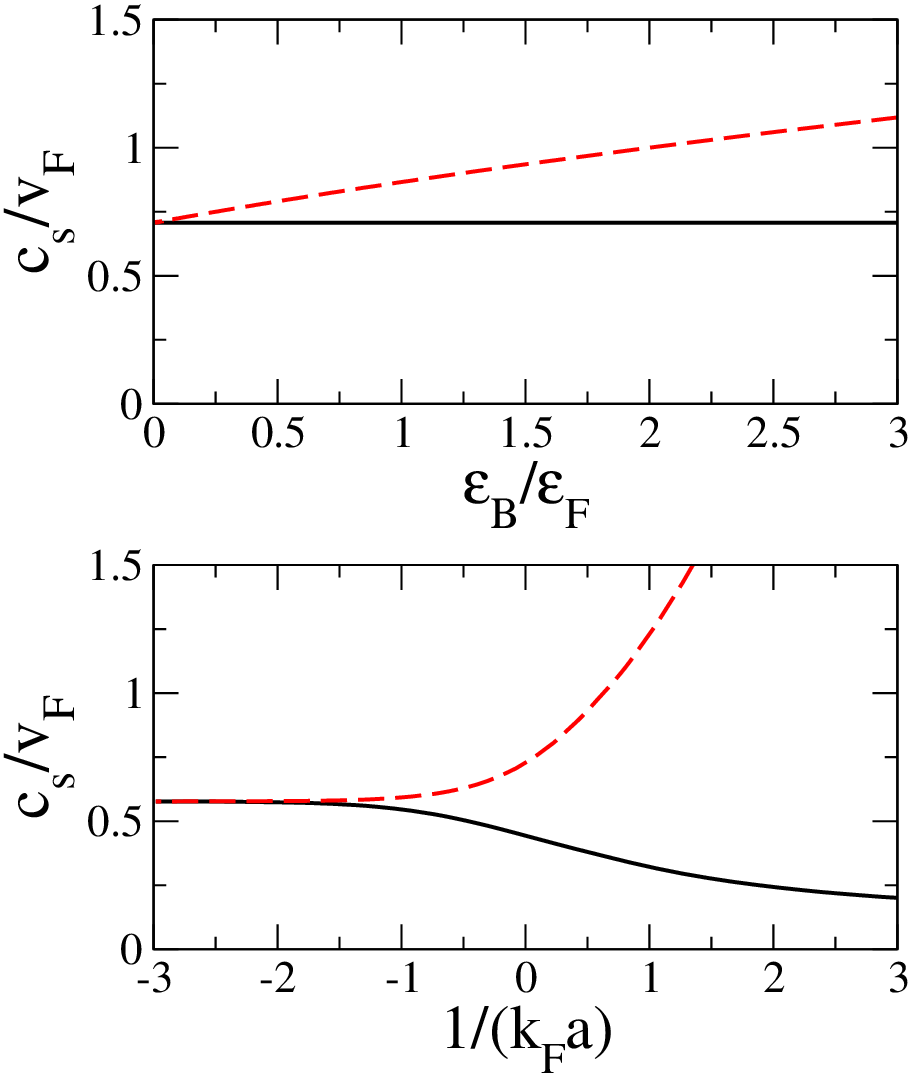,width=7cm,clip=}}
{Fig 1. Velocity $c_s$ of the Goldstone mode 
considering only phase fluctuations (dashed lines) or 
both phase and amplitude fluctuations (solid lines) of the pairing field. 
All results obtained by using the mean-field number equation 
(Equation \ref{number-mf}). \textbf{Upper panel}: 2D BCS--BEC crossover of the 
scaled velocity $c_{s}/v_F$ as a function of the scaled 
binding energy $\epsilon_B/\epsilon_F$ of the 2D Fermi superfluid. 
\textbf{Lower panel}: 3D BCS--BEC crossover of the 
scaled velocity $c_{s}/v_F$ as a function of the scaled 
inverse interaction strength $1/(k_Fa)$ of the 3D Fermi superfluid 
with scattering length $a$. 
Here $\epsilon_F=\hbar^2k_F^2/(2m)$ is the Fermi energy and 
$v_F=\sqrt{2\epsilon_F/2}$ the Fermi velocity. 
Adapted from \cite{sala2013}. BCS: Bardeen--Cooper--Schrieffer; 
BEC: Bose--Einstein condensate.}
\label{fig1}
\end{figure}   

\subsection{Connection with the Popov's Hydrodynamic Action Functional}

Given the Goldstone--Higgs action functional (\ref{low-action}) 
and performing functional integration over the Higgs 
field $\sigma({\bf r},t)$, one obtains the Goldstone action 
\beq 
S_{\theta} = \int dt \int_V d^D{\bf r} \left\{  
- {J\over 2} (\nabla \theta)^2 
+ {K\over 2} \left({\partial \theta \over \partial t}\right)^2 
\right\} 
\eeq
whose Euler--Lagrange equation is exactly Equation (\ref{euler}) 
with $K$ given by Equation (\ref{kappa}). Remarkably, this Goldstone 
action functional can be immediately derived from the 
Popov's hydrodynamic action~\cite{popov}
\beq 
S_{\theta\rho} = \int dt \int_V d^D{\bf r} \left\{  
- \hbar {\partial \theta \over \partial t} \ \rho 
- {J\over 2} (\nabla \theta)^2 
- {\hbar^2\over 2 K} \ \rho^2 
\right\}
\label{action-popov}
\eeq
functional integrating over the field $\rho({\bf r},t)$ of density 
fluctuations. $\rho({\bf r},t)$ represents a small space-time-dependent 
perturbation with respect to the constant and uniform density $n$. 
In the zero-temperature 
action functional (\ref{action-popov}), both $J$ and $K$ depend 
on $n$ through the relationship between $\mu$ and $n$. 
From the Euler--Lagrange equations of (\ref{action-popov}) with respect 
to $\theta({\bf r},t)$ and $\rho({\bf r},t)$, and introducing 
the velocity field 
\beq 
{\bf v}({\bf r},t) = {\hbar\over 2m} 
\nabla {\boldsymbol\theta}({\bf r},t) \; , 
\eeq
one finds the familiar linearized hydrodynamic equations of Euler 
\beqa 
{\partial \rho\over \partial t} 
+ n \ {\boldsymbol \nabla} \cdot {\bf v} = 0 \; , 
\label{eu1}
\\
{\partial {\bf v}\over \partial t} + {c_s^2\over n} 
{\boldsymbol \nabla} \rho = 0 \; ,  
\label{eu2}
\eeqa
from which one immediately finds the d'Alambert equation 
for density fluctuations 
\beq  
\left[ {\partial^2 \over \partial t^2} - c_s^2 \ 
\nabla^2\right] \rho = 0 \; .   
\label{euler2}
\eeq

Thus, one can identify the velocity $c_s$ of propagation of 
the Goldstone mode with the velocity $c_1$ of the first sound of the fermionic 
superfluid \cite{nagaosa,altland,stoof}. In fact, 
according to the two-fluid theory of Laudau~\cite{landau}, 
a superfluid is characterized by the presence of the first sound (where 
superfluid and normal components oscillate in phase), but also 
the second sound, where superfluid and normal components oscillate 
with opposite phases. For the sake of completeness, we observe that 
the Euler Equations (\ref{eu1}) and (\ref{eu2}) can also be 
re-written in terms of a nonlinear Schr\"odinger 
equation for the complex field $\Psi({\bf r},t) = \rho({\bf r},t)^{1/2} 
e^{i\theta({\bf r},t)}$ \cite{rivers2011,rivers2013,sala-toigo,sala-adhikari}. 

\subsection{First Sound Velocity from Thermodynamics}

On the basis of the two-fluid theory \cite{landau,khala}, 
at zero temperature the first sound velocity $c_1$ of a superfluid is 
simply given by
\beq 
c_1 = \sqrt{{n\over m} 
\left({\partial \mu\over \partial n} \right)_{V}} \; , 
\label{vero-suono}
\eeq
where $\mu$ is the chemical potential, $n$ is the number density, and $V$ 
is the volume. Refs. \cite{sala2013,combescot} have shown that 
the velocity $c_s$ of the Goldstone mode given by Equation 
(\ref{suono}) coincides with the first sound velocity $c_1$, given by 
Equation (\ref{vero-suono}), if one uses the mean-field 
equation of state (\ref{number-mf}). 

At zero temperature, the difference between Equations (\ref{number}) 
and  (\ref{number-mf}) is small in the 3D case, 
but it is instead very large in the 2D case, and in particular 
in the BEC regime of the crossover \cite{sala2016}. 
This effect is clearly shown in Figure 2, where we report 
the zero-temperature 
first sound velocity $c_1$ in the 2D BCS--BEC crossover, regulated by the 
binding energy $\epsilon_B$ of Cooper pairs. In Figure 2,
the dot-dashed line is obtained by using the mean-field number 
Equation (\ref{number-mf}), while the solid line is based on 
the beyond-mean-field number Equation (\ref{number}). 
In the strong coupling BEC regime, the beyond-mean-field equation of state 
is needed to accurately  describe the thermodynamic quantities, and the 
corresponding first found velocity $c_1$ correctly goes 
to the correct composite boson limit \cite{sala2015}. 

\begin{figure}[H]
\centerline{\epsfig{file=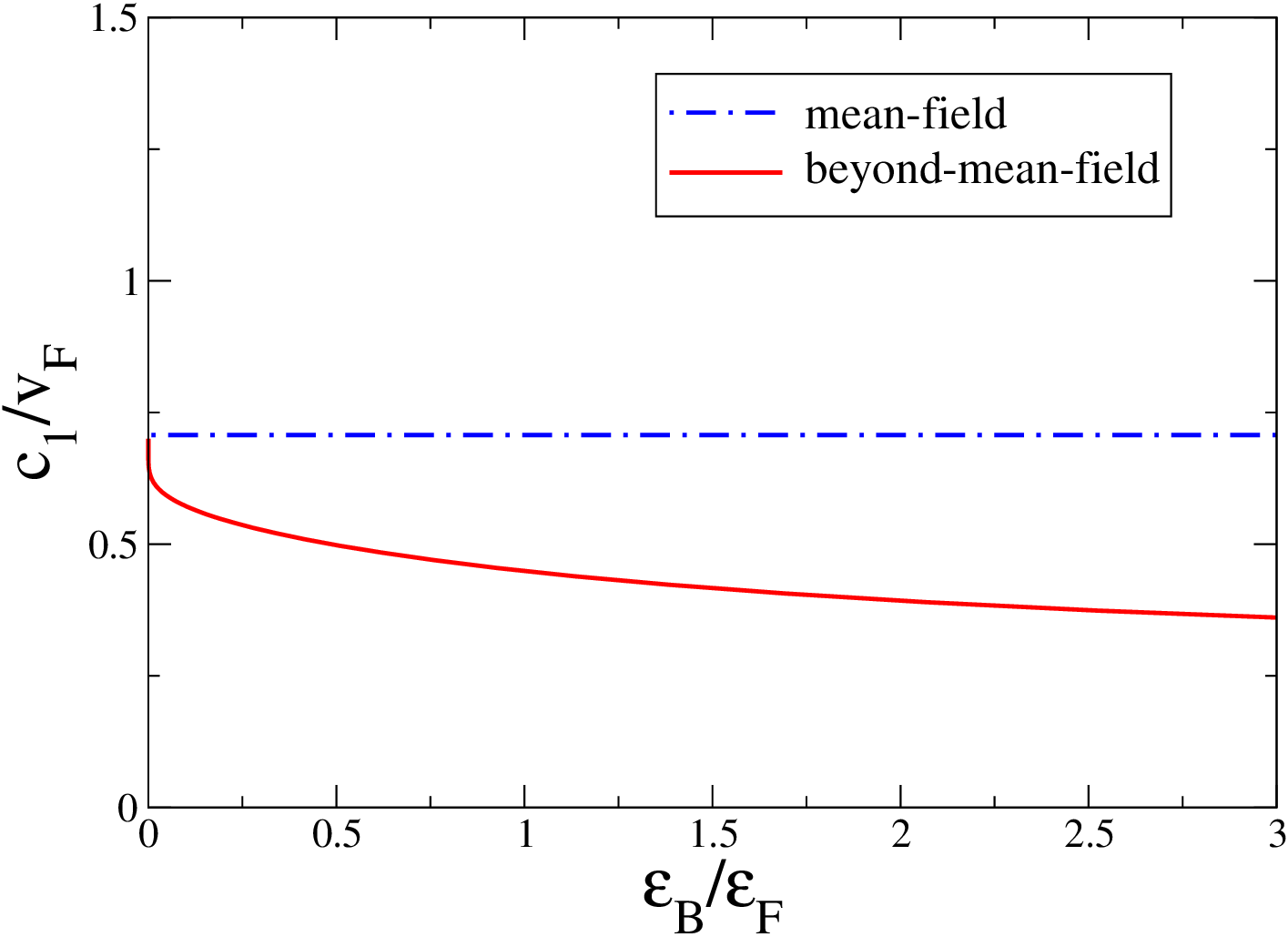,width=8cm,clip=}}
{Fig. 2. First sound velocity $c_1$ in the 2D BCS--BEC 
crossover at zero temperature, calculated by using Equation (\ref{vero-suono}). 
Here $\epsilon_B$ of the binding energy of the 
2D Fermi superfluid and $\epsilon_F=\hbar^2 \pi n/m$ is the 2D Fermi energy. 
Dot-dashed line: obtained by using the 
mean-field number equation, Equation (\ref{number-mf}). 
Solid line: obtained by using the 
beyond-mean-field number equation, Equation (\ref{number}).}
\label{fig2}
\end{figure}   

\vspace{-6pt}

{Bighin and Salasnich \cite{sala2016}} have shown that the first sound velocity 
$c_1$ obtained with the beyond-mean-field Gaussian theory 
(solid line of Figure 2) is in good agreement with very 
preliminary experimental data of $^6$Li atoms \cite{luick}. 
Unfortunately, there are not yet fully reliable published experimental 
data on the sound velocity $c_1$. 
However, there are reliable experimental data of the pressure $P$ 
at very low temperature for the dilute gas of $^6$Li 
atoms \cite{makhalov}. In Figure 3, we plot these experimental 
data (filled~circles with error bars) and compare them with our 
beyond-mean-field theory at zero temperature (solid line). 
The pressure is immediately obtained as 
\beq 
P = - {\Omega \over V} 
\eeq
with $\Omega$ given by Equation (\ref{god-or-odd}). 
Figure 3 clearly shows that, at zero temperature, 
the agreement between 
experimental data and our beyond-mean-field theory 
is very good in the full BCS--BEC crossover. In~the deep BEC 
regime, the beyond-mean-field pressure becomes \cite{sala2015} 
\beq 
P = {m\over \pi \hbar^2} \ (\mu +{1\over 2}\epsilon_B )^2 \ 
\ln{\left( {\epsilon_B \over 2(\mu + {1\over 2} \epsilon_B )} \right) } \; .
\label{eq:pbos}
\eeq

This formula (dashed line of Figure 3 is nothing else 
than the Popov equation of state \cite{popov1972} of weakly-interacting 
repulsive bosons 
\beq 
P = {m_B\over 8\pi \hbar^2} \mu_B^2 \ \ln{\left({\epsilon_B \over \mu_B} 
\right)} \; , 
\label{pressure}
\eeq
where $m_B=2m$ is the mass of composite bosons and 
$\mu_B=2 (\mu +\epsilon_B/2)$ is the bosonic chemical potential. 
Note that very recently we have obtained non-universal corrections 
to the Popov equation of state, taking account of finite-range effects 
of the inter-atomic potential \cite{sala-prl}. 

\vspace{-12pt}

\begin{figure}[H]
\centerline{\epsfig{file=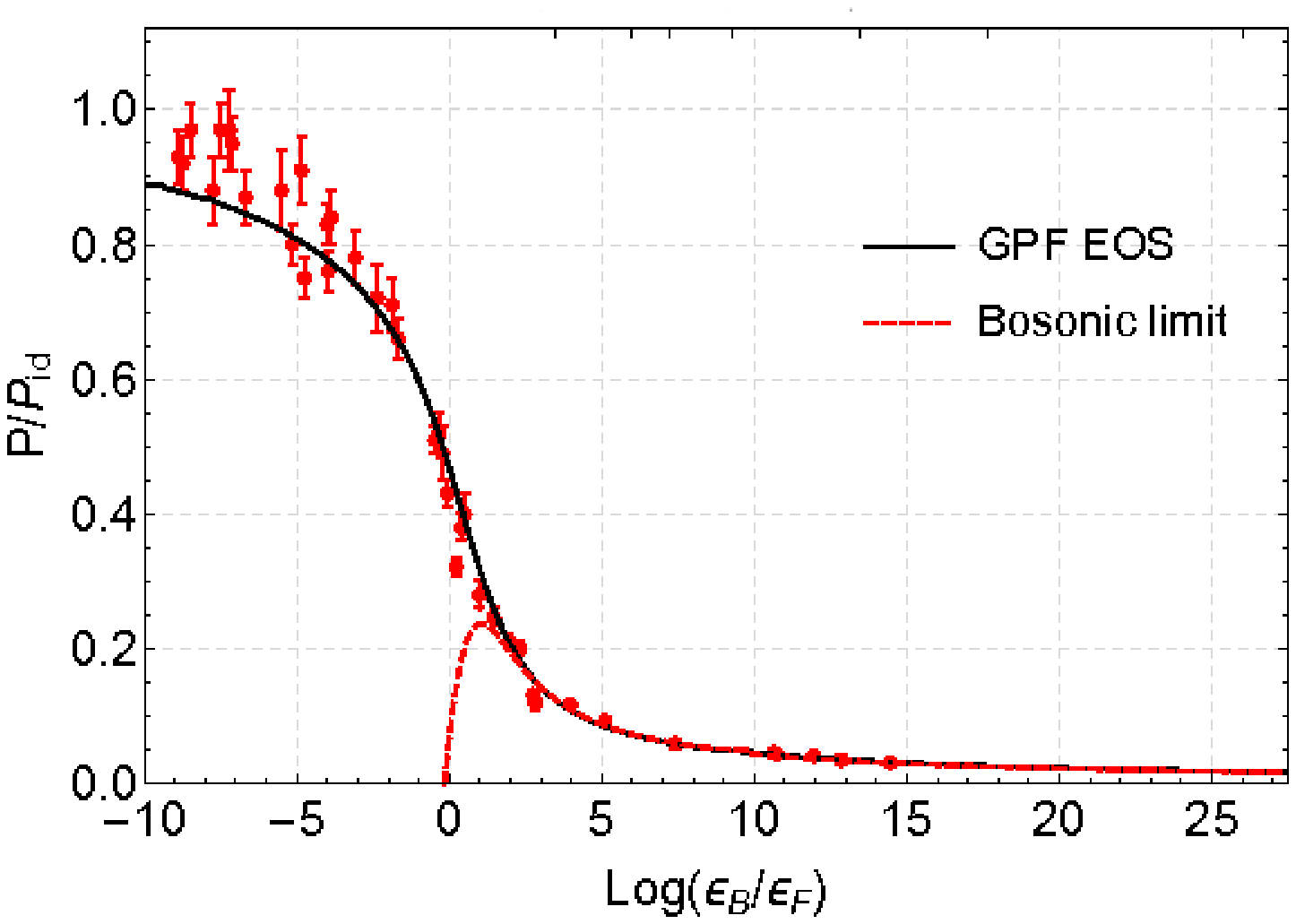,width=8cm,clip=}}
{Fig. 3. {Adimensional pressure} $P/P_{id}$ in the 2D BCS--BEC crossover, 
calculated using Gaussian pair fluctuation (GPF) 
theory (solid line) and using the Popov bosonic theory (dashed line). 
Experimental data (filled circles with error bars) {are taken 
from \cite{makhalov}.} $P_{id}$ is the pressure 
of the ideal 2D Fermi gas.}
\label{fig3}
\end{figure}   

\section{Conclusions}

We have analyzed the derivation of a low-energy 
effective field theory of Goldstone and Higgs fields 
from the beyond-mean-field BCS theory of attractive fermions. 
We have shown that across the BCS--BEC crossover, 
the inclusion of the Goldstone--Higgs 
coupling is crucial to identify the velocity of the Goldstone 
mode with the first sound velocity one gets from 
the mean-field equation of state. 
Moreover, we have explicitly shown that in the BEC side of the 
2D BCS--BEC crossover, the first sound velocity obtained 
from the beyond-mean-field 
equation of state is quite different with respect to the mean-field one. 
Finally, comparing our theoretical results with 2D experimental data, 
we have found that at zero temperature the beyond-mean-field theory  
based on Gaussian pair fluctuations seems reliable in the full 
2D BCS--BEC crossover, giving the Popov equation of state 
in the deep BEC regime. However, as shown in 
Refs. \cite{giancarlo,jacques}, at low temperatures and 
in the deep BCS regime, quantized vortices and dark solitons 
obtained with effective field approaches based on 
low-energy expansion may contradict with the Bogoliubov-de Gennes 
theory, which is well-founded in the BCS regime.
\vspace{6pt}

\conflictsofinterest{The author declares no conflict of interest.}


\begin{thebibliography}{999}

\bibitem{greiner2003} Greiner, M.; Regal, C.A.; Jin, D.S. 
Emergence of a Molecular Bose-Einstein Condensate 
from a Fermi Gas. \emph{Nature} \textbf{2003}, \emph{426}, 537--540. 

\bibitem{chin2004} Chin, C.; Bartenstein, M; Altmeyer, A; Riedl, S.; 
Jochim, S.; Hecker Denschlag, J.; Grimm, R. 
Observation of the Pairing Gap in a Strongly 
Interacting Fermi Gas. \emph{Science} \textbf{2004}, \emph{305}, 1128--1130. 

\bibitem{smerzi} Smerzi, A.; Fantoni, S.; Giovanazzi, S.; Shenoy, S.R. 
Quantum coherent atomic tunneling between two trapped 
Bose-Einstein condensates. \emph{Phys. Rev. Lett}. \textbf{1997}, \emph{79}, 
4950. 

\bibitem{sala1999} Salasnich, L.; Parola, A.; Reatto, L. 
Bose condensate in a double-well trap: 
Ground state and elementary excitations. \emph{Phys. Rev. A} \textbf{1999}, 
\emph{60}, 4171--4174. 

\bibitem{sala-mqt} Salasnich, L.; Manini, N.; Toigo, F. 
Macroscopic periodic tunneling of Fermi atoms in the BCS-BEC crossover. 
\emph{Phys. Rev. A} \textbf{2008}, \emph{77}, 043609. 

\bibitem{roati} Valtolina, G.; Burchianti, A.; Amico, A.; Neri, E.; 
Xhani, K.; Seman, J.A.; Trombettoni, A.; Smerzi, A.; Zaccanti, A.; 
Inguscio, M.; Roati, G. Josephson effect in fermionic superfluids 
across the BEC-BCS crossover. 
\emph{Science} \textbf{2016}, \emph{350}, 1505--1508. 

\bibitem{makhalov} Makhalov, V.; Martiyanov, K.; Turlapov, A. 
Ground-state pressure of quasi-2D Fermi and Bose gases. 
\emph{Phys.~Rev. Lett.}  \textbf{2014}, \emph{112}, 045301.

\bibitem{murthy} Murthy, P.A.; Boettcher, I.; Bayha, L.; 
Holzmann, M.; Kedar, D.; Neidig, M.; Ries, M.G.; Wenz, A.N.; 
Zurn, G.; Jochim, S. Observation of the Berezinskii-Kosterlitz-Thouless 
phase transition in an ultracold Fermi gas. 
\emph{Phys. Rev. Lett.} \textbf{2015}, \emph{115}, 010401.

\bibitem{fenech} 
Fenech, K.; Dyke, P.; Peppler, T.; Lingham, M.G.; 
Hoinka, S.; Hu, H.; Vale, C.J. 
Thermodynamics of an attractive 2D Fermi gas. 
\emph{Phys. Rev. Lett.} \textbf{2016}, \emph{116}, 045302.

\bibitem{boettcher} 
Boettcher, I.; Bayha, L.; Kedar, D.; Murthy, P.A.; Neidig, M.; Ries, M.G.; 
Wenz, A.N.; Zurn, G.; Jochim, S.; Enss, T. 
Equation of state of ultracold fermions in the 2D BEC-BCS crossover. 
\emph{Phys. Rev. Lett.}  \textbf{2016},~\emph{116}, 045303.

\bibitem{sala2013}
Salasnich, L.; Marchetti, P.A.; Toigo, F. 
Superfluidity, sound velocity, and quasicondensation 
in the two-dimensional BCSBEC crossover. 
\emph{Phys. Rev. A} \textbf{2013}, \emph{88}, 053612. 

\bibitem{sala2015}
Salasnich, L.; Toigo, F. 
Composite bosons in the two-dimensional BCS-BEC crossover from 
Gaussian fluctuations. \emph{Phys. Rev. A} \textbf{2015}, \emph{91}, 011604. 

\bibitem{cin2015}
He, L.; Lv, H.; Cao, G.; Hu, H.; Liu, X.-J.
Quantum fluctuations in the BCS-BEC crossover of two-dimensional 
Fermi gases. \emph{Phys. Rev. A} \textbf{2015}, \emph{92}, 023620. 

\bibitem{cin2015b} Mulkerin, B.C.; Fenech, K.; Dyke, P.; Vale, C.J.; Liu, 
X.-J.; Hu, H. 
Comparison of strong-coupling theories for a two-dimensional Fermi gas. 
\emph{Phys. Rev. A} \textbf{2015}, \emph{92}, 063636. 

\bibitem{sala2016} 
Bighin, G.; Salasnich, L. 
Finite-temperature quantum fluctuations in two-dimensional 
Fermi superfluids. \emph{Phys. Rev. B} \textbf{2016}, \emph{93}, 014519. 

\bibitem{sala-review}
Salasnich, L.; Toigo, F. 
Zero-point energy of ultracold atoms. 
\emph{Phys. Rep.} \textbf{2016}, \emph{640}, 1--29. 

\bibitem{cin2017} 
Mulkerin, B.C.;  {et al}. Superfluid Density and Critical Velocity 
Near the Fermionic Berezinskii- Kosterlitz-Thouless Transition, 
e-preprint arXiv:1702.07091.

\bibitem{larkin-varlamov}
Larkin, A.; Varlamov, A. \emph{Theory of Fluctuations in Superconductors}; 
Oxford University Press: Oxford, UK, 2005. 

\bibitem{levin}
Chen, Q.; Stajic, J.; Tan, S.; Levin, K. 
BCS–BEC crossover: From high temperature superconductors
to ultracold superfluids. \emph{Phys. Rep.} \textbf{2005}, \emph{412}, 1--88. 

\bibitem{okazaji} Okazaji, K.;  {et al.} 
Superconductivity in an electron band just above the Fermi level. 
\emph{Sci. Rep.} \textbf{2014}, \emph{4}, 4109. 

\bibitem{berezinskii} Berezinskii, V.L. Destruction of long-range order in 
one-dimensional and two-dimensional systems possessing a continuous symmetry 
group. II. Quantum systems. \textit{Sov. Phys. JETP} \textbf{1972}, 
\emph{34}, 610--616.

\bibitem{kosterlitz} Kosterlitz, J.M.; Thouless, D.J. Ordering, 
metastability and phase transitions in two-dimensional systems. 
\textit{J.~Phys. C Solid State Phys.} \textbf{1973}, \emph{6}, 1181--1203.

\bibitem{sala2017} 
Bighin, G.; Salasnich, L. 
Vortices and antivortices in two-dimensional ultracold Fermi gases. 
\emph{Sci. Rep.} \textbf{2017},~\emph{7},~45702. 

\bibitem{nelson} Nelson, D.R.; Kosterlitz, J.M. Universal 
jump in the superfluid density of two-dimensional superfluids. 
\textit{Phys.~Rev. Lett.} \textbf{1977}, \emph{39}, 1201--1205.

\bibitem{nagaosa} Nagaosa, N. 
{\it Quantum Field Theory in Condensed Matter Physics}; 
Springer: Berlin, Germany, 1999. 

\bibitem{altland} Altland, A.; Simons, B. 
{\it Condensed Matter Field Theory}; 
Cambridge University Press: Cambridge, UK, 2006. 

\bibitem{stoof} Stoof, H.T.C.; Gubbels, K.B.; Dickerscheid, 
D.B.M. {\it Ultracold Quantum Fields}; 
Springer: Dordrecht, The~Netherlands, 2009. 

\bibitem{varma} Pekker, D.; Varma, C.M. 
Amplitude/Higgs modes in condensed matter physics. 
\emph{Annu. Rev. Condens. Matter~Phys.} \textbf{2015}, \emph{6}, 269--297. 

\bibitem{Tempere:2012cm}
Tempere, J.; Devreese, J.P.A. {\it Superconductors---Materials, Properties
and Applications}; InTech: Rijeka, Croatia, 2012; pp. 1--32. 

\bibitem{griffin} Taylor, E.; Griffin, A.; Fukushima, 
N.; Ohashi, Y. 
Pairing fluctuations and the superfluid density 
through the BCS-BEC crossover. 
\emph{Phys. Rev. A} \textbf{2006}, \emph{74}, 063626.
\bibitem{drumm} Hu, H.; Liu, X.-J.; Drummond, P.D. 
Equation of state of a superfluid Fermi gas in the BCS-BEC. 
\emph{Europhys.~Lett}. \textbf{2006}, \emph{74}, 574. 

\bibitem{nsr} Nozieres, P.; Schmitt-Rink, S. 
Bose condensation in an attractive fermion gas: 
From weak to strong coupling superconductivity. 
\emph{J. Low Temp. Phys.} \textbf{1985}, \emph{59}, 195--211. 

\bibitem{moit} Engelbrecht, J.R.; Randeria, M.; Sa de Melo, C.A.R. 
BCS to Bose crossover: Broken-symmetry state. 
\emph{Phys.~Rev.~B} \textbf{1997}, \emph{55}, 15153--15156. 

\bibitem{moit2} Diener, R.B.; Sensarma, R.; Randeria, M. 
Quantum Fluctuations in the Superfluid State of the BCS-BEC Crossover. 
\emph{Phys. Rev. A} \textbf{2008}, \emph{77}, 023626. 

\bibitem{schakel} Schakel, A.M.J. 
Derivation of the effective action of a dilute Fermi gas 
in the unitary limit of the BCS-BEC crossover. 
\emph{Ann. Phys}. \textbf{2011}, \emph{326}, 193--206. 

\bibitem{rivers2011} Lin, C.-Y.; Lee, D.-S.; Rivers, R.J. 
Spontaneous vortex production in driven condensates with narrow 
Feschbach resonances. \emph{Phys. Rev. A} \textbf{2011}, \emph{84}, 013623. 

\bibitem{rivers2013} Hsiang, J.-T.; Lin, 
C.-Y.; Lee, D.-S.; Rivers, R.J. 
The role of causality in tunable Fermi gas condensates. 
\emph{J. Phys. Condens. Matter} \textbf{2013}, \emph{25}, 404211. 

\bibitem{levin2} Anderson, B.M.; Boyack, R.; Wu, C.-T.; Levin, K. 
Correcting inconsistencies in the conventional superfluid 
path integral scheme. \emph{Phys. Rev. B} \textbf{2016}, \emph{93}, 180504. 

\bibitem{popov}  Popov, V.N. {\it Functional Integrals 
in Quantum Field Theory and Statistical Physics}; 
Reidel: Dordrecht, The~Netherlands, 1983.

\bibitem{landau}  {Landau, L.D. The theory of superfuidity of helium II. 
\emph{Phys. Rev.}  \textbf{1941}, \emph{60}, 356--358.}

\bibitem{sala-toigo} Salasnich, L.; Toigo, F. 
Extended Thomas-Fermi density functional for the unitary Fermi gas. 
\emph{Phys. Rev. A} \textbf{2008}, \emph{78}, 053626. 

\bibitem{sala-adhikari} Adhikari, S.K.; Salasnich, L. 
Effective nonlinear Schrodinger equations for cigar-shaped 
and disc-shaped Fermi superfluids at unitarity. \emph{New J. Phys}. 
\textbf{2009}, \emph{11}, 023011. 

\bibitem{khala} Khalatnikov, I.M. {\it An Introduction to the Theory of 
Superfluidity}; Avalon Publishing: New York, NY, USA, 1965.

\bibitem{combescot} Combescot, R.; Kagan, M.Y.; Stringari, S. 
Collective mode of homogeneous superfluid Fermi gases in the 
BEC-BCS crossover. \emph{Phys. Rev. A} \textbf{2006}, \emph{74}, 042717. 

\bibitem{luick} Luick, N. (Local Probing of the Berezinskii-Kosterlitz-Thouless 
Transition in a Two-Dimensional Bose Gas. Master Thesis, 
University of Hamburg). Unpublished work, 2014. 

\bibitem{popov1972} Popov, V.N. On the theory of the superfluidity 
of two- and one-dimensional Bose systems. 
\emph{Theor. Math. Phys.~A} \textbf{1972}, \emph{11}, 565--573.

\bibitem{sala-prl} Salasnich, L. 
Non-Universal equation of state of the two-dimensional Bose gas. 
\emph{Phys. Rev. Lett.} \textbf{2017}, \emph{118},~130402. 

\bibitem{giancarlo} Simonucci, S.; Strinati, G.C. 
Equation for the superfluid gap obtained by coarse graining 
the Bogoliubov---De~Gennes equations throughout the BCS-BEC crossover. 
\emph{Phys. Rev. B} \textbf{2014}, \emph{89}, 054511. 

\bibitem{jacques} Lombardi, G.; Van Alphen, W.; Klimin, S.N.; Tempere, 
J. Soliton-core filling in superfluid Fermi gases 
with spin imbalance. \emph{Phys. Rev. A} \textbf{2016}, \emph{93}, 013614. 

\end{thebibliography}
\end{document}